\documentclass[12pt]{iopart}

\usepackage{graphicx}
\usepackage{epsfig}
\usepackage[latin1]{inputenc}
\usepackage{amssymb}

\begin{document}

\title[]{Dilepton production as a measure of QGP thermalization time}

\author{Mauricio Martinez$^{1}$ and Michael Strickland$^2$}
\address{$^1$ Helmholtz Research School, Johann Wolfgang Goethe Universit\"at Frankfurt, \\
Ruth-Moufang-Str. 1, 60438 Frankfurt am Main, Germany\\
$^2$ Institut f\"{u}r Theoretische Physik, Johann Wolfgang Goethe 
Universit\"{a}t,\\
Max von Laue-Str. 1, 60438 Frankfurt am Main, Germany\\[0.2ex]
}

\begin{abstract}
 We discuss dilepton production from a quark-gluon plasma which has a time-dependent anisotropy in momentum space. A phenomenological model for the hard momentum scale, $p_{\rm hard}(\tau)$ and the plasma anisotropy parameter, $\xi(\tau)$, is presented. The model interpolates between early-time $1+1$ free streaming behavior ($\tau \ll \tau_{\rm iso}$) and late-time ideal $1+1$ hydrodynamical behavior ($\tau \gg \tau_{\rm iso}$). Using this model, we find that the dilepton rate arising from electromagnetic annihilations of quarks in the kinematic range 3  $<p_T<$  8 GeV is sensitive to the assumed isotropization time of the system, $\tau_{iso}$. Therefore high-energy dilepton production can be used to probe the degree of momentum-space isotropy of a quark-gluon plasma produced in relativistic heavy ion collisions and the time of onset of hydrodynamic expansion of the QGP.
\end{abstract}

\pacs{25.75.Nq,25.75.Dw,12.38.Mh,11.10.Wx.}


\section{Introduction}

One of the most interesting problems facing the community in relativistic heavy ion collisions is to determine at what time the matter created can be described using hydrodynamics. In this context, at RHIC energies it has been found that for $p_T \lesssim 2$ GeV, the elliptic flow of the matter created is described well by models which assume ideal hydrodynamic behavior starting at very early times $\tau \lesssim$ $1$ fm/c ~\cite{Teaney:2000cw,Huovinen:2001cy,Hirano:2002ds,Tannenbaum:2006ch}. This is not completely understood due to the fact that the estimates from perturbative QCD for the thermalization time of a QGP at RHIC energies range from $2-3$ fm/c~\cite{Baier:2000sb,Xu:2004mz,Strickland:2007fm}. In the hydrodynamical description of the matter generated during heavy ion collistions, there is a strong dependence on the equation of state, initial conditions, late time evolution ($\gtrsim$ 2-3 fm/c), etc. It would be nice to have information about the thermalization time and appropiate initial conditions from other independent observables which are different than elliptic flow. In this work, we examine the possibility to determine experimentally  the thermalization time of the matter created using high energy dilepton yields as a function of both, mass and transverse momentum. We compute the expected $e^+e^-$ yields resulting from a Pb-Pb collision at LHC full beam energy, $\sqrt{s} = 5.5$ TeV in a pre-equilibrium scenario of a quark-gluon plasma with a time-dependent anisotropy caused by the rapid longitudinal expansion. 

\section{Dilepton production}

\begin{figure}[t]
\includegraphics[width=15.0cm]{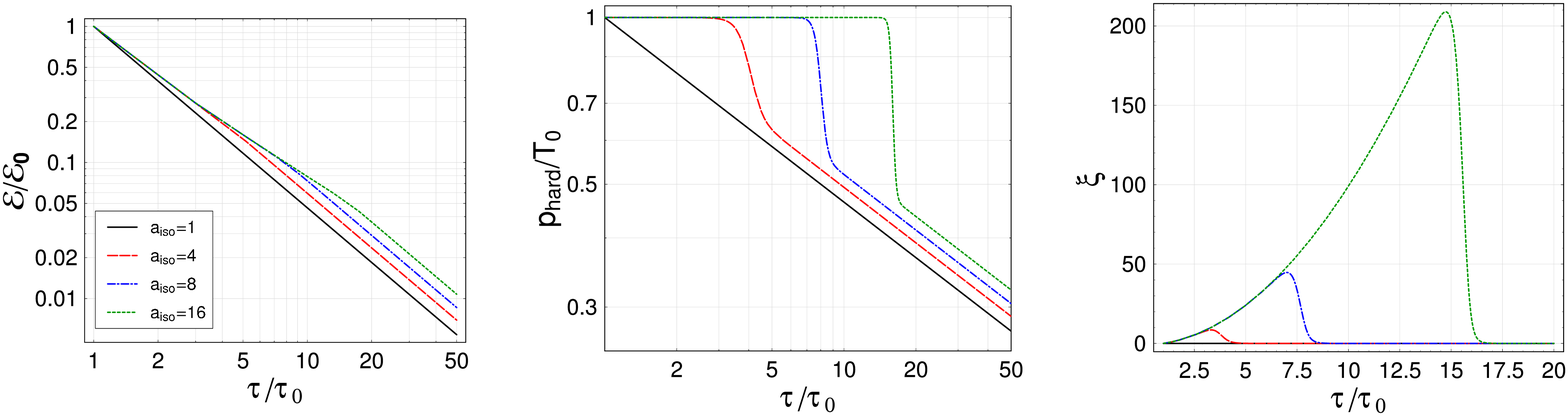}
\vspace{-2mm}
\caption{Model energy density (left), hard momentum scale (middle),
and anisotropy parameter (right) for four
different isotropization times $\tau_{\rm iso} \in \{0.1,0.4,0.8,1.6\}$ fm/c 
assuming $\tau_0 =$ 0.1 fm/c. 
The transition width is taken to be $\gamma = 2$.}
\label{fig:modelPlot}
\end{figure}


From relativistic kinetic theory, the dilepton production rate at leading order is given by:
\begin{equation}
E \frac{d R}{d^3P} = \int \frac{d^3{\bf p}_1}{(2\pi)^3}\,\frac{d^3{\bf p}_2}{(2\pi)^3}\,f_q({\bf p}_1)\,f_{\bar{q}}({\bf p}_2)\, \it{v}_{q\bar{q}}\,\sigma^{l^+l^-}_{q\bar{q}}\,\delta^{(4)}(P-p_1-p_2)
        \; ,
\label{eq:annihilation1}
\end{equation}
where $f$ is the distribution function in phase space for the quark (antiquark),$\it{v}_{q\bar{q}}$ is the relative velocity between the quark and antiquark and $\sigma^{l^+l^-}_{q\bar{q}}$ is the total cross section at leading order in $\alpha_{em}$ for the reaction $q\bar{q} \to l^{+} l^{-}$. We will assume that the anisotropic phase distribution can be obtained from an arbitrary isotropic phase space distribution changing its argument, i.e., $f_i({\bf p},\xi,p_{\rm hard})=f_{iso}^i(\sqrt{{\bf p^2}+\xi({\bf p\cdot \hat{e}_z})^{\bf 2}},p_{\rm hard})$, here $i={q,\bar{q}}$, $p_{\rm hard}$ is the hard momentum scale, $\hat{e}_z$ is the direction of the anisotropy and $\xi >0$ is a parameter that reflects the strength and type of anisotropy. In equilibrium, $\xi$=0 and $p_{\rm hard}$ can be identified with the plasma temperature $T$.

\subsection{Interpolating Model}
For any anisotropic phase space distribution $f_i({\bf p},\xi,p_{\rm hard})$, the medium parton energy density can be factorized as:
\begin{equation}
{\cal E}(p_{\rm hard},\xi) \, = \, 
    \int \frac{d^3{\bf p}}{2\pi^3} \; p \, f({\bf p},\xi) \, = \, 
    {\cal E}_{0}(p_{\rm hard}) \; {\cal R}(\xi) \;\, ,
\end{equation}
where ${\cal R}(\xi) = \left[ 1/(\xi+1) + {\rm 
arctan}\sqrt{\xi}/\sqrt{\xi} \right]/2$ and ${\cal 
E}_0(p_{\rm hard})$ is the energy density resulting from integration 
of the isotropic quark and anti-quark distribution 
functions.
In order to construct a model which interpolates between free streaming 
and hydrodynamic expansion we introduce a smeared step function 
$\lambda(\tau,\tau_0,\tau_{\rm iso},\gamma) \equiv \left({\rm 
tanh}\left[\gamma (\tau-\tau_{\rm iso})/\tau_0 \right]+1\right)/2$.  This allows
us to model the time-dependence of $p_{\rm hard}$ and $\xi$ as \cite{Mauricio:2007vz}:

\begin{eqnarray}
{\cal E}(\tau) &=& {\cal E}_{\rm FS}(\tau) \,
                    \left[\,{\cal U}(\tau)/{\cal U}(\tau_0)\,\right]^{4/3} \; , \nonumber \\
p_{\rm hard}(\tau) &=& T_0 \, \left[\,{\cal U}(\tau)/{\cal U}(\tau_0)\,\right]^{1/3} \; ,  \\
\xi(\tau) &=& a^{2(1-\lambda(\tau))} - 1 \; , \nonumber
\label{eq:modelEQs}
\end{eqnarray}
where ${\cal U}(\tau) \equiv \left[{\cal R}\!\left(a_{\rm iso}^2-
1\right)\right]^{3\lambda(\tau)/4}\left(a_{\rm 
iso}/a\right)^{\lambda(\tau)}$, $a \equiv \tau/\tau_0$ and $a_{\rm 
iso} \equiv \tau_{\rm iso}/\tau_0$. When $\tau \ll \tau_{\rm iso}$ we have $\lambda \rightarrow 0$ and the system is free streaming.  When $\tau \gg 
\tau_{\rm iso}$ then $\lambda \rightarrow 1$ and the system is 
expanding hydrodynamically.  In the limit $\gamma \rightarrow 
\infty$, $\lambda \rightarrow \Theta(\tau-\tau_{\rm iso})$.  In 
Fig.~\ref{fig:modelPlot} we plot the time-dependence of ${\cal E}$, 
$p_{\rm hard}$, and $\xi$ assuming $\gamma=2$ for four different plasma 
isotropization times corresponding to $a_{\rm iso} \in \{1,4,8,16\}$.

To obtain the final expected dilepton yields we integrate the 
production rate over $\tau \in \{\tau_0,\tau_f\}$ and 
$\eta \in \{-2.5,2.5\}$ with parameters specified by Eq.~(\ref{eq:modelEQs}) and $\tau_f$ set by $p_{\rm hard}(\tau_f)=T_c$: 
\begin{eqnarray}
  \frac{dN}{dM^2dy}&=\pi R^2_T\int d^2P_T\int_{\tau_0}^{\tau_f}\int_{-\infty}^{\infty}\frac{dR}{d^4P}\tau d\tau d\eta\hspace{0.2cm}, \label{Mspectrum}\\
        \frac{dN}{d^2P_Tdy}&=\pi R^2_T\int dM^2\int_{\tau_0}^{\tau_f}\int_{-\infty}^{\infty}\frac{dR_{\rm ann}}{d^4P}\tau d\tau d\eta\hspace{0.2cm}.
\label{pTspectrum}
\end{eqnarray}
where $R_T\,=\,1.2\,A^{1/3}$ fm is the radius of the nucleus in the transverse plane. 

\section{Results}

\begin{figure}
  \begin{center}
  \includegraphics[width=7.25cm,height=5.5cm]{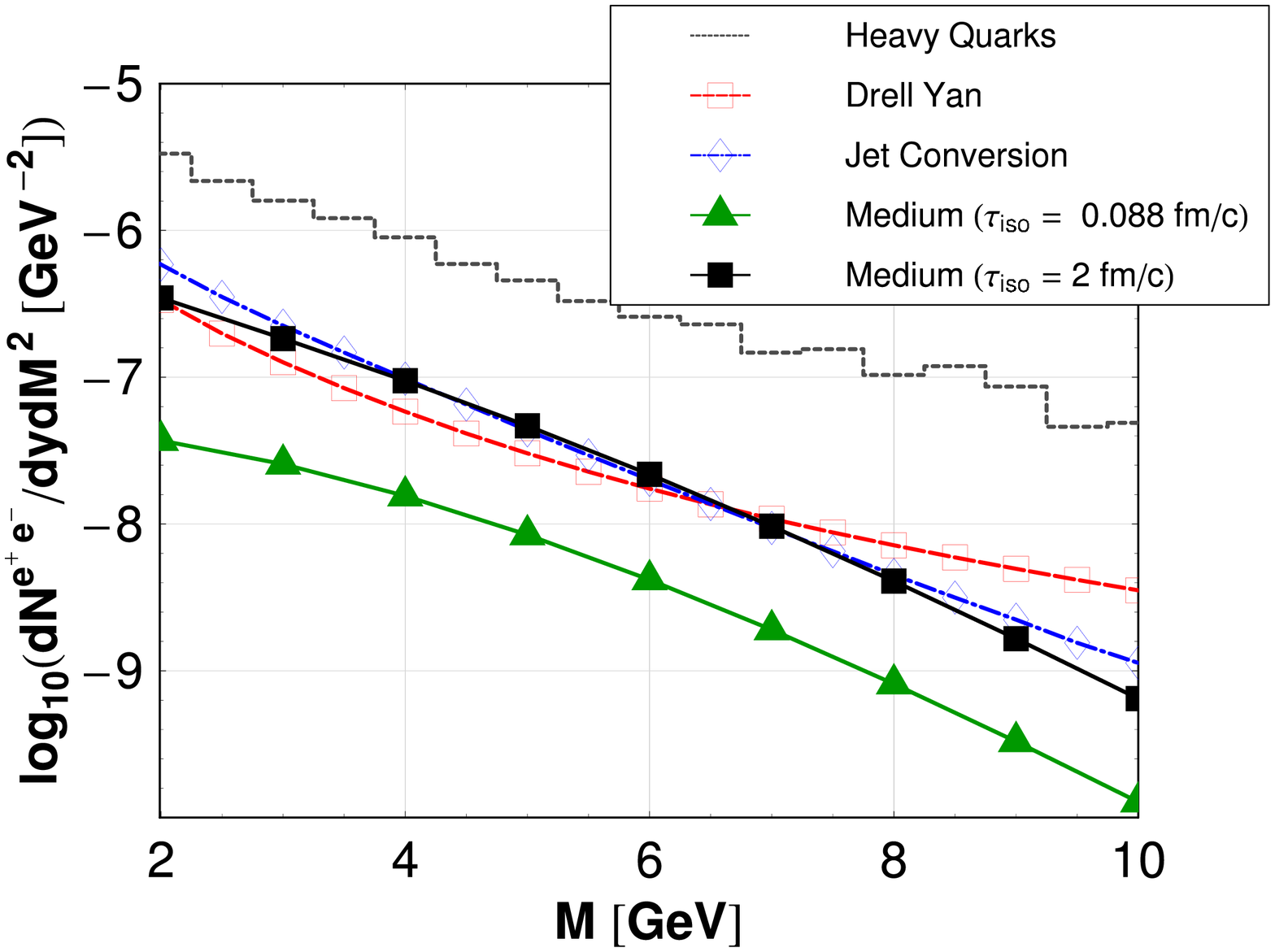}
\hspace{0.7cm}
  \includegraphics[width=7.45cm,height=5.5cm]{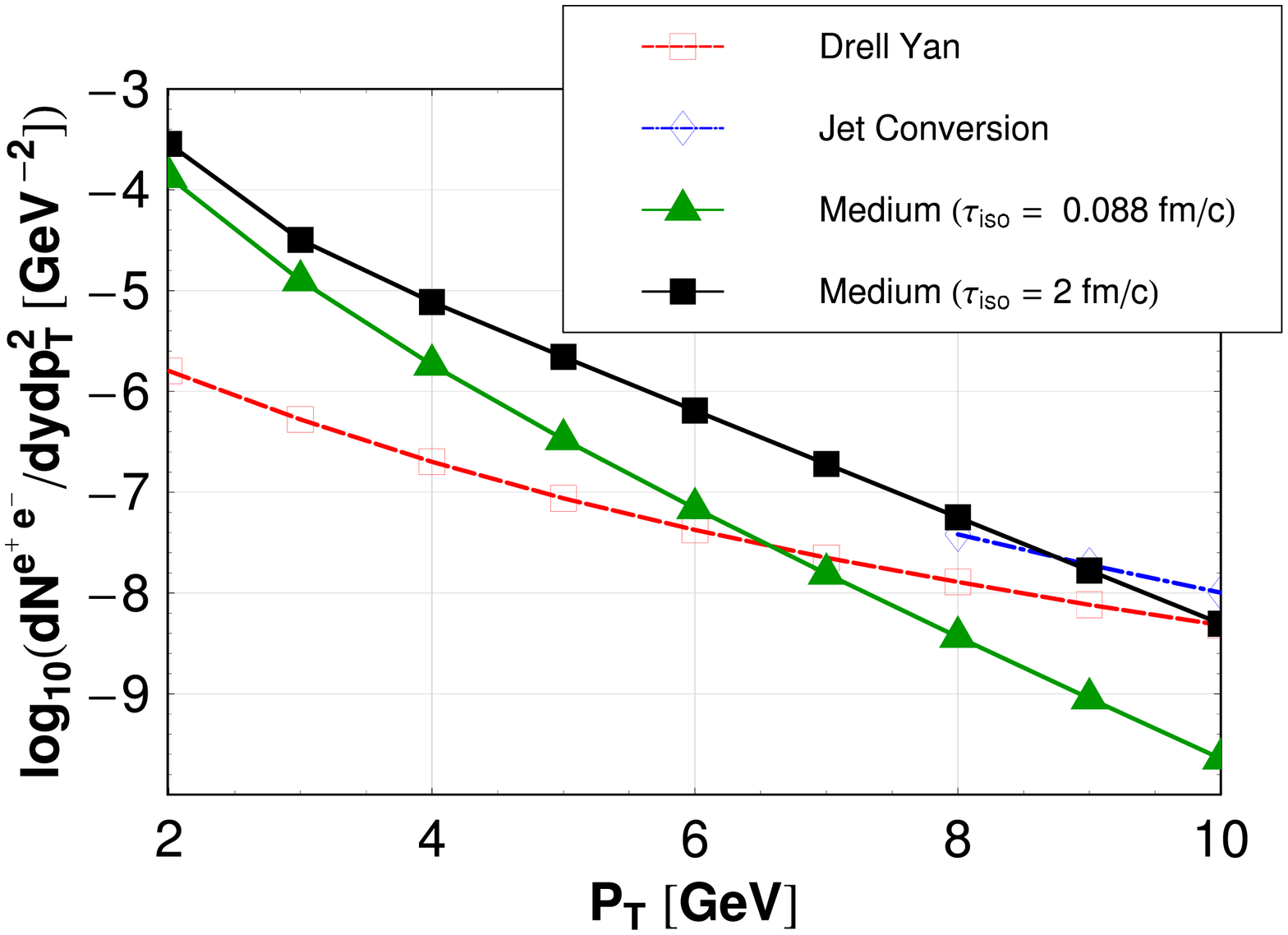}
\end{center}
\vspace{-2mm}
\caption{\label{dilep} (Color online) Invariant mass (left) and momentum (right) distribution of midrapidity dileptons in central Pb+Pb collisions at LHC.}
\end{figure}

 In Fig.~\ref{dilep} we show our final predicted $e^+e^-$ yields as a function of invariant mass and transverse momentum along with predicted yields from other sources. For comparison with previous works we take $\tau_0 = 0.088$ fm/c, $T_0 = 845$ MeV, $T_c = 160$ MeV, and $R_T = 7.1$ fm~\cite{Turbide:2006mc}. Here we assume that when the system reaches 
$T_c$ all medium emission stops. The addition of mixed and hadronic 
phase emission is not included in the present work since those effects are not important in the studied kinematic region~\cite{Mauricio:2007vz}. 
As a function of the invariant mass (left side of Fig.~\ref{dilep}), the in-medium production is sensitive when varying the assumed plasma isotropization time from $0.088$ fm/c to 2 fm/c.  When $\tau_{\rm iso}\sim$ 2 fm/c, we see that medium dileptons become as important as Drell-Yan and jet conversion. However, all three contributions are down by an order of magnitude from the expected background coming from semileptonic heavy quark decay.
As a function of $p_T$ (right side of Fig.~\ref{dilep}), the medium contribution dominates the expected Drell-Yan and jet conversion sources for all $p_T \lesssim 6$ GeV.  If $\tau_{\rm iso}\sim$ 2 fm/c then the in-medium dileptons dominate out to $p_T \sim 9$ GeV.  In the same figure, at $p_T =$ 5 GeV the expected medium dilepton yield varies by nearly an order of magnitude depending on the assumed plasma isotropization time.

\section{Conclusion}

Based on the results from Fig.~\ref{dilep}, dilepton production as a function of the transverse momentum is a good observable to measure $\tau_{\rm iso}$ at LHC energies in the kinematic range $3 < p_T < 8$ GeV. Additionally, it may be possible to estimate the maximum amount of momentum-space anisotropy achieved 
during the lifetime of the QGP using the phenomenological model presented here. 
The effect of varying $\tau_{\rm iso}$ is also large in the dilepton 
spectra vs invariant mass but Drell-Yan and jet conversion production can be up to 10 times larger than medium production making it difficult to measure a clean medium dilepton signal.
Future work in dilepton production incorporating anisotropies will study the effect of including collisional broadening of the parton distributions, the 
possibility of late-time persistent anisotropies (finite viscosity), NLO order corrections and finite chemical potentials.
In addition, models such as (\ref{eq:modelEQs}) can be used to assess
the impact of momentum-space anisotropies on other observables such as jet conversion.

\ack
We thank A.~Dumitru, B.~Schenke, S.~Turbide and A. Ipp. M.M. was supported by the Helmholtz Research School. M.S. was supported by DFG project GR 1536/6-1.

\end{document}